\documentclass[12pt,a4paper]{article}
\usepackage{amsmath}
\usepackage{latexsym}
\usepackage{amssymb}
\usepackage{graphicx,color}
\makeatletter
\def\rddots{\mathinner{\mkern1mu\raise\p@%
    \vbox{\kern7\p@\hbox{.}}\mkern2mu%
    \raise4\p@\hbox{.}\mkern2mu\raise7\p@\hbox{.}\mkern1mu}}
\makeatother
\makeatletter
\def\eqnarray{%
\stepcounter{equation}%
\let\@currentlabel=\theequation
\global\@eqnswtrue
\global\@eqcnt\z@
\tabskip\@centering
\let\\=\@eqncr
$$\halign to \displaywidth\bgroup\@eqnsel\hskip\@centering
$\displaystyle\tabskip\z@{##}$&\global\@eqcnt\@ne
\hfil$\displaystyle{{}##{}}$\hfil
&\global\@eqcnt\tw@$\displaystyle\tabskip\z@{##}$\hfil
\tabskip\@centering&\llap{##}\tabskip\z@\cr}
\makeatother
\newcommand{\ket}[1]{{\vert{#1}\rangle}}

\newcommand{\fukuso}{{\mathbf C}}

\begin{document}

\title{\sl A Master Equation with Generalized Lindblad Form and 
a Unitary Transformation by the Squeezing Operator}
\author{
  Kazuyuki FUJII
  \thanks{E-mail address : fujii@yokohama-cu.ac.jp }\\
  ${}^{\dagger}$Department of Mathematical Sciences\\
  Yokohama City University\\
  Yokohama, 236--0027\\
  Japan
  }
\date{}
\maketitle
%
%
%
%
\begin{abstract}
  In the preceding paper arXiv : 0802.3252 [quant-ph] we treated 
  a model given by a master equation with generalized Lindblad form, 
  and examined the algebraic structure related to some Lie algebras 
  and constructed an approximate solution.
  
  In this paper we apply a unitary transformation by the squeezing 
  operator to the master equation. Then the generalized Lindblad form 
  is tranformed to the usual Lindblad one, while the (original) 
  Hamiltonian is tranformed to somewhat complicated one. 

  As a result we have two different representations based on the 
  Lie algebra $su(1,1)$. We examine new algebraic structure and 
  construct some approximate solution.
\end{abstract}
%


%
%
%
%

Quantum Computation (Computer) is one of main subjects in Quantum 
Physics. To realize it we must overcome severe problems arising from 
Decoherence, so we need to study Quantum Open System to control  
decoherence (if possible).

This paper is a series of \cite{Fujii}, \cite{EFS} and \cite{Fujii-2} 
and we continue to study dynamics of a quantum open system.  
First we explain our purpose in a short manner. See \cite{BP} as a 
general introduction to this subject.
 
We consider a quantum open system $S$ coupled to the environment $E$. 
Then the total system $S+E$ is described by the Hamiltonian
\[
H_{S+E}=H_{S}\otimes {\bf 1}_{E}+{\bf 1}_{S}\otimes H_{E}+H_{I}
\]
where $H_{S}$, $H_{E}$ are respectively the Hamiltonians of the system and 
environment, and $H_{I}$ is the Hamiltonian of the interaction.

\par \noindent
Then under several assumptions (see \cite{BP}) the reduced dynamics of the 
system (which is not unitary !) is given by the Master Equation
\begin{equation}
\label{eq:master-equation}
\frac{\partial}{\partial t}\rho=-i[H_{S},\rho]-{\cal D}(\rho)
\end{equation}
with the dissipator being the usual Lindblad form
\begin{equation}
\label{eq:dissipator}
{\cal D}(\rho)=\frac{1}{2}\sum_{\{j\}}
\left(A_{j}^{\dagger}A_{j}\rho+\rho A_{j}^{\dagger}A_{j}
-2A_{j}\rho A_{j}^{\dagger}\right).
\end{equation}
Here $\rho\equiv \rho(t)$ is the density operator (or matrix) of the system. 

Similarly, the master equation of quantum damped harmonic oscillator 
(see \cite{BP}, Section 3.4.6 or also \cite{WS}) is given by
\begin{equation}
\label{eq:quantum damped harmonic oscillator}
\frac{\partial}{\partial t}\rho=-i[\omega a^{\dagger}a,\rho]
-
\frac{\mu}{2}
\left(a^{\dagger}a\rho+\rho a^{\dagger}a-2a\rho a^{\dagger}\right)
-
\frac{\nu}{2}
\left(aa^{\dagger}\rho+\rho aa^{\dagger}-2a^{\dagger}\rho{a}\right),
\end{equation}
where $a$ and $a^{\dagger}$ are the annihilation and creation operators of 
the system (for example, an electro--magnetic field mode in a cavity), and 
$\mu,\ \nu$ ($\mu > \nu \geq 0$) are some real constants depending on 
the system (for example, a damping rate of the cavity mode).

In \cite{EFS} the general solution was given in {\bf the operator algebra 
level}. This is a very important step. 
Moreover, in \cite{Fujii} we treated the master equation with 
generalized Lindblad (or Kossakowski--Lindblad \cite{ABF}) form given by
\begin{eqnarray}
\label{eq:quantum damped harmonic oscillator generalized}
\frac{\partial}{\partial t}\rho
=-i[\omega a^{\dagger}a,\rho]
&&-\frac{\mu}{2}
\left(a^{\dagger}a\rho+\rho a^{\dagger}a-2a\rho a^{\dagger}\right)
-\frac{\nu}{2}
\left(aa^{\dagger}\rho+\rho aa^{\dagger}-2a^{\dagger}\rho{a}\right)
\nonumber \\
&&-\frac{\kappa}{2}\left(a^{2}\rho+\rho a^{2}-2a\rho a\right)
  -\frac{\bar{\kappa}}{2}\left( (a^{\dagger})^{2}\rho+
   \rho (a^{\dagger})^{2}-2a^{\dagger}\rho a^{\dagger}\right)
\end{eqnarray}
where $\kappa$ is a complex constant satisfying the condition 
${\mu}{\nu}\geq |\kappa|^{2}$ which ensures the positivity. 

\par \noindent
Then we examined the algebraic structure related to the Lie algebras 
$su(1,1)$ and $su(2)$, and constructed some approximate solutions 
by use of it.

In this paper we apply a unitary transformation by the squeezing 
operator to (\ref{eq:quantum damped harmonic oscillator generalized}), 
which was developed by An et al. \cite{AWL}, \cite{AWLJ} \footnote{
In the papers the positivity condition ${\mu}{\nu}\geq |\kappa|^{2}$ 
has been neglected}. 
Then the generalized Lindblad form is tranformed to the usual 
Lindblad one, while the (original) Hamiltonian is tranformed to 
somewhat complicated one. We examine new algebraic structure in detail. 

The squeezing operator $S(\epsilon)$ is given by
\begin{equation}
S(\epsilon)=\exp\left(\frac{1}{2}\left(\epsilon(a^{\dagger})^{2}-
\bar{\epsilon}a^{2}\right)\right),\quad 
\epsilon=|\epsilon|e^{i\phi} \in \fukuso
\end{equation}
($|\epsilon|$ and $\phi$ are chosen later on) and it has well--known 
remarkable property
\begin{equation}
S(\epsilon)aS(\epsilon)^{\dagger}=
c(|\epsilon|)a-e^{i\phi}s(|\epsilon|)a^{\dagger},
\quad 
S(\epsilon)a^{\dagger}S(\epsilon)^{\dagger}=
c(|\epsilon|)a^{\dagger}-e^{-i\phi}s(|\epsilon|)a
\end{equation}
where $c(|\epsilon|)\equiv \cosh(|\epsilon|)$ and 
$s(|\epsilon|)\equiv \sinh(|\epsilon|)$ for simplicity, see 
for example \cite{KF1}.

We act the squeezing operator to the master equation 
(\ref{eq:quantum damped harmonic oscillator generalized}) as  
adjoint action
\begin{eqnarray}
\label{eq:quantum damped harmonic oscillator generalized-modified}
\frac{\partial}{\partial t}\rho_{S}
&=&-i[\omega {a_{S}}^{\dagger}a_{S},\rho_{S}] \nonumber \\
&&
-\frac{\mu}{2}
\left({a_{S}}^{\dagger}a_{S}\rho_{S}+
\rho_{S}{a_{S}}^{\dagger}a_{S}-2a_{S}\rho_{S}{a_{S}}^{\dagger}\right)
-\frac{\nu}{2}
\left(a_{S}{a_{S}}^{\dagger}\rho_{S}+
\rho_{S} a_{S}{a_{S}}^{\dagger}-2{a_{S}}^{\dagger}\rho_{S}a_{S}\right)
\nonumber \\
&&
-\frac{\kappa}{2}\left({a_{S}}^{2}\rho_{S}+\rho_{S} {a_{S}}^{2}
 -2a_{S}\rho_{S} a_{S}\right)
-\frac{\bar{\kappa}}{2}\left( ({a_{S}}^{\dagger})^{2}\rho_{S}+
 \rho_{S} ({a_{S}}^{\dagger})^{2}-2{a_{S}}^{\dagger}\rho_{S} 
 {a_{S}}^{\dagger}\right)
\end{eqnarray}
where $\rho_{S}=S\rho S^{\dagger}$ and $a_{S}=Sa S^{\dagger}$. 
Now let us calculate the right hand side. 

\par \noindent
It is not difficult to check
\begin{eqnarray}
\label{eq:calculation-1}
A_{S}
&\equiv& 
{a_{S}}^{\dagger}a_{S}\rho_{S}+\rho_{S}{a_{S}}^{\dagger}a_{S}
-2a_{S}\rho_{S}{a_{S}}^{\dagger}  \nonumber \\
&=&
c^{2}\left\{a^{\dagger}a\rho_{S}+\rho_{S}a^{\dagger}a
-2a\rho_{S}a^{\dagger}\right\}
+
s^{2}\left\{aa^{\dagger}\rho_{S}+\rho_{S}aa^{\dagger}
-2a^{\dagger}\rho_{S}a\right\} \nonumber \\
&&-
e^{-i\phi}cs\left\{a^{2}\rho_{S}+\rho_{S}a^{2}-2a\rho_{S}a\right\}
-e^{i\phi}cs\left\{(a^{\dagger})^{2}\rho_{S}+\rho_{S}(a^{\dagger})^{2}
                   -2a^{\dagger}\rho_{S}a^{\dagger}\right\}
\end{eqnarray}
and
\begin{eqnarray}
\label{eq:calculation-2}
B_{S}
&\equiv& 
a_{S}{a_{S}}^{\dagger}\rho_{S}+\rho_{S}a_{S}{a_{S}}^{\dagger}
-2{a_{S}}^{\dagger}\rho_{S}a_{S}  \nonumber \\
&=&
s^{2}\left\{a^{\dagger}a\rho_{S}+\rho_{S}a^{\dagger}a
-2a\rho_{S}a^{\dagger}\right\}
+
c^{2}\left\{aa^{\dagger}\rho_{S}+\rho_{S}aa^{\dagger}
-2a^{\dagger}\rho_{S}a\right\} \nonumber \\
&&-
e^{-i\phi}cs\left\{a^{2}\rho_{S}+\rho_{S}a^{2}-2a\rho_{S}a\right\}
-e^{i\phi}cs\left\{(a^{\dagger})^{2}\rho_{S}+\rho_{S}(a^{\dagger})^{2}
                   -2a^{\dagger}\rho_{S}a^{\dagger}\right\}
\end{eqnarray}
and
\begin{eqnarray}
\label{eq:calculation-3}
C_{S}
&\equiv& 
a_{S}^{2}\rho_{S}+\rho_{S}a_{S}^{2}-2{a_{S}}\rho_{S}a_{S}  \nonumber \\
&=&
-e^{i\phi}cs\left\{a^{\dagger}a\rho_{S}+\rho_{S}a^{\dagger}a
-2a\rho_{S}a^{\dagger}\right\}
-e^{i\phi}cs\left\{aa^{\dagger}\rho_{S}+\rho_{S}aa^{\dagger}
-2a^{\dagger}\rho_{S}a\right\} \nonumber \\
&&+
c^{2}\left\{a^{2}\rho_{S}+\rho_{S}a^{2}-2a\rho_{S}a\right\}
+e^{2i\phi}s^{2}\left\{(a^{\dagger})^{2}\rho_{S}+\rho_{S}(a^{\dagger})^{2}
                   -2a^{\dagger}\rho_{S}a^{\dagger}\right\}
\end{eqnarray}
and
\begin{eqnarray}
\label{eq:calculation-4}
D_{S}
&\equiv& 
({a_{S}}^{\dagger})^{2}\rho_{S}+\rho_{S}({a_{S}}^{\dagger})^{2}
-2{a_{S}}^{\dagger}\rho_{S}{a_{S}}^{\dagger}  \nonumber \\
&=&
-e^{-i\phi}cs\left\{a^{\dagger}a\rho_{S}+\rho_{S}a^{\dagger}a
-2a\rho_{S}a^{\dagger}\right\}
-e^{-i\phi}cs\left\{aa^{\dagger}\rho_{S}+\rho_{S}aa^{\dagger}
-2a^{\dagger}\rho_{S}a\right\} \nonumber \\
&&+
e^{-2i\phi}s^{2}\left\{a^{2}\rho_{S}+\rho_{S}a^{2}-2a\rho_{S}a\right\}
+c^{2}\left\{(a^{\dagger})^{2}\rho_{S}+\rho_{S}(a^{\dagger})^{2}
                   -2a^{\dagger}\rho_{S}a^{\dagger}\right\}
\end{eqnarray}
where $c=c(|\epsilon|)$ and $s=s(|\epsilon|)$ for simplicity.

Therefore by (\ref{eq:calculation-1}) $\sim$ (\ref{eq:calculation-4}) 
the generalized Lindblad form in 
(\ref{eq:quantum damped harmonic oscillator generalized-modified}) 
becomes 
\begin{eqnarray}
\label{eq:Lindblad-f}
&&-\frac{1}{2}
\left(
\mu c^{2}+\nu s^{2}-\kappa e^{i\phi}cs-\bar{\kappa}e^{-i\phi}cs
\right)A
-\frac{1}{2}
\left(
\mu s^{2}+\nu c^{2}-\kappa e^{i\phi}cs-\bar{\kappa}e^{-i\phi}cs
\right)B \\
\label{eq:Lindblad-l}
&&-\frac{1}{2}
\left(
-\mu e^{-i\phi}cs-\nu e^{-i\phi}cs+\kappa c^{2}+e^{-2i\phi}\bar{\kappa}s^{2}
\right)C
-\frac{1}{2}
\left(
-\mu e^{i\phi}cs-\nu e^{i\phi}cs+\bar{\kappa}c^{2}+e^{2i\phi}{\kappa}s^{2}
\right)D \nonumber \\
&&
\end{eqnarray}
where $A=A_{S={\bf 1}}$ et al. 

Now we can remove the terms (\ref{eq:Lindblad-l}) if we choose $\epsilon$ 
suitably. In fact, we set
\[
\kappa=|\kappa|e^{-i\phi}\equiv ke^{-i\phi}
\]
then we have a quadratic equation
\begin{equation}
\label{eq:quadratic equation}
-(\mu+\nu)cs+k(c^{2}+s^{2})=0
\Longleftrightarrow 
kt^{2}-(\mu+\nu)t+k=0
\end{equation}
for $t\equiv \frac{s}{c}=\tanh(|\epsilon|)$. The discriminant 
${\cal D}$ is
\[
{\cal D}=(\mu+\nu)^{2}-4k^{2}=(\mu-\nu)^{2}+4(\mu\nu-k^{2})\geq 0
\]
by the positivity condition $\mu\nu \geq k^{2}$, so there is a 
solution on $t\ (\Longrightarrow on\ |\epsilon|)$.

{\bf A comment is in order}. We cannot remove the second term in  
(\ref{eq:Lindblad-f}) against the result in \cite{AWLJ}, \cite{AWL} 
by the positivity condition. We leave it to readers.

\vspace{3mm} \noindent
Therefore, the generalized Lindblad form becomes
\begin{equation}
\label{eq:G-Lindblad}
-\frac{\mu c^{2}+\nu s^{2}-2kcs}{2}
\left(a^{\dagger}a\rho_{S}+\rho_{S}a^{\dagger}a-2a\rho_{S}a^{\dagger}\right)
-\frac{\mu s^{2}+\nu c^{2}-2kcs}{2}
\left(aa^{\dagger}\rho_{S}+\rho_{S}aa^{\dagger}-2a^{\dagger}\rho_{S}{a}\right).
\end{equation}

Next let us calculate the Hamiltonian in 
(\ref{eq:quantum damped harmonic oscillator generalized-modified}), 
which is easy to see
\begin{equation}
-i[\omega {a_{S}}^{\dagger}a_{S},\rho_{S}]
=
-i[\omega\left\{(c^{2}+s^{2})a^{\dagger}a
-e^{i\phi}cs(a^{\dagger})^{2}-e^{-i\phi}cs a^{2}\right\},\rho_{S}].
\end{equation}

As a result the master equation 
(\ref{eq:quantum damped harmonic oscillator generalized-modified}) 
becomes
\begin{eqnarray}
\label{eq:final master equation}
\frac{\partial}{\partial t}\rho_{S}
&=&
-i[\omega\left\{(c^{2}+s^{2})a^{\dagger}a
-e^{i\phi}cs(a^{\dagger})^{2}-e^{-i\phi}cs a^{2}\right\},\rho_{S}]
\nonumber \\
&&
-\frac{\mu c^{2}+\nu s^{2}-2kcs}{2}
\left(a^{\dagger}a\rho_{S}+\rho_{S}a^{\dagger}a-2a\rho_{S}a^{\dagger}\right)
\nonumber \\
&&
-\frac{\mu s^{2}+\nu c^{2}-2kcs}{2}
\left(aa^{\dagger}\rho_{S}+\rho_{S}aa^{\dagger}-2a^{\dagger}\rho_{S}{a}\right)
\end{eqnarray}
under the squeezing operator $S(\epsilon)$ with some 
$\epsilon=|\epsilon|e^{i\phi}$.

Next we apply the method in \cite{Fujii}, \cite{EFS} to the equation 
(\ref{eq:final master equation}) to make the algebraic structure 
clearer. Before that let us make some mathematical preparations. 

A matrix representation of $a$ and $a^{\dagger}$ on the usual Fock 
space
\[
{\cal F}=\mbox{Vect}_{\fukuso}\{\ket{0},\ket{1},\ket{2},\ket{3},\cdots \};
\quad \ket{n}=\frac{(a^{\dagger})^{n}}{\sqrt{n!}}\ket{0}
\]
is given by

\begin{eqnarray}
\label{eq:creation-annihilation}
a&=&\mbox{e}^{i\theta}
\left(
\begin{array}{ccccc}
0 & 1 &          &          &        \\
  & 0 & \sqrt{2} &          &        \\
  &   & 0        & \sqrt{3} &        \\
  &   &          & 0        & \ddots \\
  &   &          &          & \ddots
\end{array}
\right),\quad
a^{\dagger}=\mbox{e}^{-i\theta}
\left(
\begin{array}{ccccc}
0 &          &          &        &        \\
1 & 0        &          &        &        \\
  & \sqrt{2} & 0        &        &        \\
  &          & \sqrt{3} & 0      &        \\
  &          &          & \ddots & \ddots
\end{array}
\right) 
\\
N&\equiv& a^{\dagger}a=
\left(
\begin{array}{ccccc}
0 &   &   &   &        \\
  & 1 &   &   &        \\
  &   & 2 &   &        \\
  &   &   & 3 &        \\
  &   &   &   & \ddots
\end{array}
\right)
\end{eqnarray}
where $\mbox{e}^{i\theta}$ is some phase. Note that $aa^{\dagger}
=a^{\dagger}a+1=N+1$.

For a matrix $X=(x_{ij})\in M({\cal F})$ 
\[X=
\left(
\begin{array}{cccc}
x_{11} & x_{12} & x_{13} & \cdots  \\
x_{21} & x_{22} & x_{23} & \cdots  \\
x_{31} & x_{32} & x_{33} & \cdots  \\
\vdots & \vdots & \vdots & \ddots
\end{array}
\right)
\]
we correspond to the vector $\widehat{X}\in 
{{\cal F}}^{\mbox{dim}_{\fukuso}{\cal F}}$ as
\begin{equation}
\label{eq:correspondence}
X=(x_{ij})\ \longrightarrow\ 
\widehat{X}=(x_{11},x_{12},x_{13},\cdots;x_{21},x_{22},x_{23},\cdots;
x_{31},x_{32},x_{33},\cdots;\cdots \cdots)^{T}
\end{equation}
where $T$ means the transpose. The following formula
\begin{equation}
\label{eq:well--known formula}
\widehat{AXB}=(A\otimes B^{T})\widehat{X}
\end{equation}
holds for $A,B,X\in M({\cal F})$.

If we set
\begin{equation}
\label{eq:K's I}
K_{3}=\frac{1}{2}(N\otimes {\bf 1}-{\bf 1}\otimes N),\
K_{+}=\frac{1}{2}\left\{(a^{\dagger})^{2}\otimes {\bf 1}-
{\bf 1}\otimes ((a^{\dagger})^{2})^{T}\right\},\
K_{-}=\frac{1}{2}\left\{a^{2}\otimes {\bf 1}-
{\bf 1}\otimes (a^{2})^{T}\right\}
\end{equation}
where $N^{T}=N$, then we have
\begin{equation}
\label{eq:K's relation I}
[K_{3},K_{+}]=K_{+},\quad [K_{3},K_{-}]=-K_{-},\quad 
[K_{+},K_{-}]=-2K_{3}.
\end{equation}

\par \noindent 
If we also set
\begin{equation}
\label{eq:K's II}
\tilde{K}_{3}=\frac{1}{2}(N\otimes {\bf 1}+{\bf 1}\otimes N+
{\bf 1}\otimes {\bf 1}),\quad
\tilde{K}_{+}=a^{\dagger}\otimes a^{T},\quad
\tilde{K}_{-}=a\otimes (a^{\dagger})^{T},
\end{equation}
we have
\begin{equation}
\label{eq:K's relation II}
[\tilde{K}_{3},\tilde{K}_{+}]=\tilde{K}_{+},\quad 
[\tilde{K}_{3},\tilde{K}_{-}]=-\tilde{K}_{-},\quad 
[\tilde{K}_{+},\tilde{K}_{-}]=-2\tilde{K}_{3}.
\end{equation}
Namely, $\{K_{3},K_{+},K_{-}\}$ and 
$\{\tilde{K}_{3},\tilde{K}_{+},\tilde{K}_{-}\}$ 
are a set of generators of $su(1,1)$ algebra.

\vspace{3mm}
Now the equation (\ref{eq:final master equation}) can be written 
as 
\begin{eqnarray}
\label{eq:final master equation-rewritten}
\frac{\partial}{\partial t}{\hat{\rho}}_{S}
&=&
\left\{
-2i\omega(c^{2}+s^{2})K_{3}+2i\omega e^{i\phi}csK_{+}+
2i\omega e^{-i\phi}csK_{-}
+\frac{\mu-\nu}{2}{\bf 1}\otimes {\bf 1}
\right. \nonumber \\
&&
\left.
-\left((\mu+\nu)(c^{2}+s^{2})-4kcs\right)\tilde{K}_{3}
+(\mu s^{2}+\nu c^{2}-2kcs)\tilde{K}_{+}
+(\mu c^{2}+\nu s^{2}-2kcs)\tilde{K}_{-}
\right\}{\hat{\rho}}_{S}   \nonumber \\
&\equiv&
\left(\frac{\mu-\nu}{2}{\bf 1}\otimes {\bf 1}+
{\cal A}+\tilde{{\cal A}}\right){\hat{\rho}}_{S}\ ; \\
{\cal A}&\equiv&
-2i\omega(c^{2}+s^{2})K_{3}+2i\omega e^{i\phi}csK_{+}+
2i\omega e^{-i\phi}csK_{-}, \\
\tilde{{\cal A}}&\equiv&
-\left((\mu+\nu)(c^{2}+s^{2})-4kcs\right)\tilde{K}_{3}
+(\mu s^{2}+\nu c^{2}-2kcs)\tilde{K}_{+}
+(\mu c^{2}+\nu s^{2}-2kcs)\tilde{K}_{-}
\end{eqnarray}
by use of (\ref{eq:well--known formula}) and generators 
(\ref{eq:K's I}), (\ref{eq:K's II}). The solution is formally 
given by
\begin{equation}
\label{eq:formal solution}
{\hat{\rho}}_{S}(t)=e^{\frac{\mu-\nu}{2}t}
e^{t({\cal A}+\tilde{{\cal A}})}{\hat{\rho}}_{S}(0).
\end{equation}

Now  let us calculate the commutators 
of $\{K_{3},K_{+},K_{-}\}$ and 
$\{\tilde{K}_{3},\tilde{K}_{+},\tilde{K}_{-}\}$ : 
\[
[K_{3},\tilde{K}_{+}]=[K_{3},\tilde{K}_{-}]=
[K_{3},\tilde{K}_{3}]=0
\]
and
\begin{eqnarray*}
&&
[{K}_{+},\tilde{K}_{+}]=
-a^{\dagger}\otimes (a^{\dagger})^{T},\quad 
[{K}_{+},\tilde{K}_{-}]=
-a^{\dagger}\otimes (a^{\dagger})^{T},\quad
[{K}_{+},\tilde{K}_{3}]=
-\frac{1}{2}\{(a^{\dagger})^{2}\otimes {\bf 1}+
{\bf 1}\otimes ((a^{\dagger})^{2})^{T}\},  \\
&&
[{K}_{-},\tilde{K}_{+}]=a\otimes a^{T},\quad
[{K}_{-},\tilde{K}_{-}]=a\otimes a^{T},\quad
[{K}_{-},\tilde{K}_{3}]=
\frac{1}{2}\{a^{2}\otimes {\bf 1}+
{\bf 1}\otimes (a^{2})^{T}\}.
\end{eqnarray*}

Since $\{K_{3},K_{+},K_{-}\}$ and 
$\{\tilde{K}_{3},\tilde{K}_{+},\tilde{K}_{-}\}$ 
don't commute it is reasonable to assume 
\begin{equation}
\label{eq:evolution operator}
{\hat{\rho}}_{S}(t)\approx e^{\frac{\mu-\nu}{2}t}
e^{t{\cal A}}e^{t\tilde{{\cal A}}}{\hat{\rho}}_{S}(0)
\end{equation}
as the first approximation. 

The disentangling formula (which is well--known) for the system 
$\{L_{3},L_{+},L_{-}\}$ based on the Lie algebra $su(1,1)$ 
is given by
\begin{equation}
\label{eq:disentangling formula}
e^{t(2a{L}_{3}+b{L}_{+}+c{L}_{-})}
=
e^{G(t)L_{+}}e^{-2\log(F(t))L_{3}}e^{E(t)L_{-}}
\end{equation}
with
\begin{eqnarray}
G(t)&=&\frac{\frac{b}{\sqrt{a^{2}-bc}}\sinh\left(t\sqrt{a^{2}-bc}\right)}
     {\cosh\left(t\sqrt{a^{2}-bc}\right)-\frac{a}{\sqrt{a^{2}-bc}}
      \sinh\left(t\sqrt{a^{2}-bc}\right)}, \nonumber \\
F(t)&=&\cosh\left(t\sqrt{a^{2}-bc}\right)-
     \frac{a}{\sqrt{a^{2}-bc}}\sinh\left(t\sqrt{a^{2}-bc}\right),
  \nonumber \\
E(t)&=&\frac{\frac{c}{\sqrt{a^{2}-bc}}\sinh\left(t\sqrt{a^{2}-bc}\right)}
     {\cosh\left(t\sqrt{a^{2}-bc}\right)-\frac{a}{\sqrt{a^{2}-bc}}
      \sinh\left(t\sqrt{a^{2}-bc}\right)}.
\end{eqnarray}
See for example \cite{KF1}, \cite{KF2}.

From this formula  we can calculate $e^{t{\cal A}}$ and 
$e^{t\tilde{{\cal A}}}$ in (\ref{eq:evolution operator}) as follows.
\begin{equation}
\label{eq:disentangling formula-I}
e^{t{\cal A}}
=
e^{G(t)K_{+}}e^{-2\log(F(t))K_{3}}e^{E(t)K_{-}}
\end{equation}
with
\begin{eqnarray}
G(t)&=&\frac{\frac{b}{\sqrt{a^{2}-bc}}\sinh\left(t\sqrt{a^{2}-bc}\right)}
     {\cosh\left(t\sqrt{a^{2}-bc}\right)-\frac{a}{\sqrt{a^{2}-bc}}
      \sinh\left(t\sqrt{a^{2}-bc}\right)}, \nonumber \\
F(t)&=&\cosh\left(t\sqrt{a^{2}-bc}\right)-
     \frac{a}{\sqrt{a^{2}-bc}}\sinh\left(t\sqrt{a^{2}-bc}\right),
  \nonumber \\
E(t)&=&\frac{\frac{c}{\sqrt{a^{2}-bc}}\sinh\left(t\sqrt{a^{2}-bc}\right)}
     {\cosh\left(t\sqrt{a^{2}-bc}\right)-\frac{a}{\sqrt{a^{2}-bc}}
      \sinh\left(t\sqrt{a^{2}-bc}\right)}
\end{eqnarray}
and
\begin{equation}
a=-i\omega(c^{2}+s^{2}),\quad 
b=2i\omega e^{i\phi}cs,\quad
c=2i\omega e^{-i\phi}cs.
\end{equation}
Then it is easy to see $\sqrt{a^{2}-bc}=i\omega$ and 
\begin{eqnarray}
\label{eq:components-1}
G(t)&=&\frac{2ie^{i\phi}cs\sin(t\omega)}
     {\cos(t\omega)+i(c^{2}+s^{2})\sin(t\omega)},\quad  
F(t)=\cos(t\omega)+i(c^{2}+s^{2})\sin(t\omega), \nonumber \\
E(t)&=&\frac{2ie^{-i\phi}cs\sin(t\omega)}
     {\cos(t\omega)+i(c^{2}+s^{2})\sin(t\omega)}.
\end{eqnarray}

Next,
\begin{equation}
\label{eq:disentangling formula-II}
e^{t\tilde{{\cal A}}}
=
e^{\tilde{G}(t)\tilde{K}_{+}}
e^{-2\log(\tilde{F}(t))\tilde{K}_{3}}
e^{\tilde{E}(t)\tilde{K}_{-}}
\end{equation}
with
\begin{eqnarray}
\label{eq:components-2}
\tilde{G}(t)
&=&\frac{\frac{b}{\sqrt{a^{2}-bc}}\sinh\left(t\sqrt{a^{2}-bc}\right)}
     {\cosh\left(t\sqrt{a^{2}-bc}\right)-\frac{a}{\sqrt{a^{2}-bc}}
      \sinh\left(t\sqrt{a^{2}-bc}\right)}, \nonumber \\
\tilde{F}(t)
&=&\cosh\left(t\sqrt{a^{2}-bc}\right)-
     \frac{a}{\sqrt{a^{2}-bc}}\sinh\left(t\sqrt{a^{2}-bc}\right),
  \nonumber \\
\tilde{E}(t)
&=&\frac{\frac{c}{\sqrt{a^{2}-bc}}\sinh\left(t\sqrt{a^{2}-bc}\right)}
     {\cosh\left(t\sqrt{a^{2}-bc}\right)-\frac{a}{\sqrt{a^{2}-bc}}
      \sinh\left(t\sqrt{a^{2}-bc}\right)}
\end{eqnarray}
and
\begin{equation}
a=-\frac{(\mu+\nu)(c^{2}+s^{2})-4kcs}{2},\quad  
b=\mu s^{2}+\nu c^{2}-2kcs,\quad
c=\mu c^{2}+\nu s^{2}-2kcs.
\end{equation}
Then it is not difficult to see
\[
\sqrt{a^{2}-bc}=\frac{\mu-\nu}{2}.
\]
However, by use of it we cannot simplify (\ref{eq:components-2})  
like (\ref{eq:components-1}).

Therefore our approximate solution 
(\ref{eq:evolution operator}) becomes
\begin{equation}
\label{eq:approximate form}
{\hat{\rho}}_{S}(t)\approx e^{\frac{\mu-\nu}{2}t}
e^{G(t)K_{+}}e^{-2\log(F(t))K_{3}}e^{E(t)K_{-}}
e^{\tilde{G}(t)\tilde{K}_{+}}
e^{-2\log(\tilde{F}(t))\tilde{K}_{3}}
e^{\tilde{E}(t)\tilde{K}_{-}}
{\hat{\rho}}_{S}(0)
\end{equation}
and we restore this form to the usual one by use of 
(\ref{eq:well--known formula}). The result is
\begin{eqnarray}
\label{eq:final form-1}
{\rho}_{S}(t)\approx 
&&
e^{\frac{\mu-\nu}{2}t}
\exp\left(\frac{G(t)}{2}(a^{\dagger})^{2}\right)
\left\{
\exp\left(-\log(F(t))N\right)\times  
\right.  \nonumber \\
&&
\left.
\left\{
\exp\left(\frac{E(t)}{2}a^{2}\right)
\phi(t)
\exp\left(-\frac{E(t)}{2}a^{2}\right)
\right\}\exp\left(\log(F(t))N\right)  
\right\}
\exp\left(-\frac{G(t)}{2}(a^{\dagger})^{2}\right)
\end{eqnarray}
and
\begin{eqnarray}
\label{eq:final form-2}
\phi(t)=
\frac{1}{\tilde{F}(t)}
&&\sum_{n=0}^{\infty}
\frac{\tilde{G}(t)^{n}}{n!}(a^{\dagger})^{n}
\left\{
\exp\left(-\log(\tilde{F}(t))N\right)\times  
\right.
\nonumber \\
&&
\left\{
\sum_{m=0}^{\infty}
\frac{\tilde{E}(t)^{m}}{m!}a^{m}{\rho}_{S}(0)(a^{\dagger})^{m}
\right\}
\left.
\exp\left(-\log(\tilde{F}(t))N\right)
\right\}
a^{n}.
\end{eqnarray}
This is indeed complicated. Compare this with the corresponding 
one in \cite{Fujii}.

\vspace{5mm}
In this paper we revisited the quantum damped harmonic oscillator 
with generalized Lindblad form and applied the unitary transformation 
by the squeezing operator to the master equation, and examined the new 
algebraic structure and next constructed some approximate solution 
in the operator algebra level. 

The model is very important to understand several phenomena related 
to quantum open systems, so the general solution is indeed required. 
To obtain it (like in \cite{EFS}) is almost impossible at the present 
time.

Lastly, we conclude the paper by stating our motivation. We are studying 
a model of quantum computation (computer) based on Cavity QED (see 
\cite{FHKW1} and \cite{FHKW2}), so in order to construct a more realistic 
model of (robust) quantum computer we have to study severe problems 
coming from decoherence. 

For example, we have to study the quantum damped Jaynes--Cummings model 
(in our terminology) whose phenomenological master equation for the 
density operator is given by
\begin{equation}
\label{eq:quantum damped Jaynes-Cummings}
\frac{\partial}{\partial t}\rho=-i[H_{JC},\rho]
-
\frac{\mu}{2}
\left(a^{\dagger}a\rho+\rho a^{\dagger}a-2a\rho a^{\dagger}\right)
-
\frac{\nu}{2}
\left(aa^{\dagger}\rho+\rho aa^{\dagger}-2a^{\dagger}\rho{a}\right),
\end{equation}
where $H_{JC}$ is the well--known Jaynes-Cummings Hamiltonian given by
\begin{eqnarray}
H_{JC}
&=&
\frac{\omega_{0}}{2}\sigma_{3}\otimes {\bf 1}+ 
\omega_{0}1_{2}\otimes a^{\dagger}a +
\Omega\left(\sigma_{+}\otimes a+\sigma_{-}\otimes a^{\dagger}\right)
\\
&=&
\left(
  \begin{array}{cc}
    \frac{\omega_{0}}{2}+\omega_{0}N & \Omega a             \\
    \Omega a^{\dagger} & -\frac{\omega_{0}}{2}+\omega_{0}N
  \end{array}
\right) \nonumber
\end{eqnarray}
with
\[
\sigma_{+} = 
\left(
  \begin{array}{cc}
    0 & 1 \\
    0 & 0
  \end{array}
\right), \quad 
\sigma_{-} = 
\left(
  \begin{array}{cc}
    0 & 0 \\
    1 & 0
  \end{array}
\right), \quad 
\sigma_{3} = 
\left(
  \begin{array}{cc}
    1 & 0  \\
    0 & -1
  \end{array}
\right), \quad 
{\bf 1}_{2} = 
\left(
  \begin{array}{cc}
    1 & 0 \\
    0 & 1
  \end{array}
\right). 
\]
Note that $\rho \in M(2;\fukuso)\otimes M({\cal F})=
M(2;M({\cal F}))$, where $M({\cal F})$ is the set of all 
operators on the Fock space ${\cal F}$. 
See for example \cite{Scala et al-1}, \cite{Scala et al-2}.

Furthermore, it may be possible to treat the generalized master 
equation given by
\begin{eqnarray}
\label{eq:quantum damped Jaynes-Cummings II}
\frac{\partial}{\partial t}\rho
&=&-i[H_{JC},\rho]
-
\frac{\mu}{2}
\left(a^{\dagger}a\rho+\rho a^{\dagger}a-2a\rho a^{\dagger}\right)
-
\frac{\nu}{2}
\left(aa^{\dagger}\rho+\rho aa^{\dagger}-2a^{\dagger}\rho{a}\right)
\nonumber \\
&&-\frac{\kappa}{2}\left(a^{2}\rho+\rho a^{2}-2a\rho a\right)
  -\frac{\bar{\kappa}}{2}\left( (a^{\dagger})^{2}\rho+
   \rho (a^{\dagger})^{2}-2a^{\dagger}\rho a^{\dagger}\right)
\end{eqnarray}
with the condition ${\mu}{\nu}\geq |\kappa|^{2}$ similarly  
in this paper.

These equations ((\ref{eq:quantum damped Jaynes-Cummings}), 
(\ref{eq:quantum damped Jaynes-Cummings II})) are very hard 
to solve in the operator algebra level,  
so even constructing approximate solutions is not easy. 
This is our future task.


\end{document}